\documentclass [12pt]{article}
\usepackage {epsfig}
\usepackage{graphicx}

\begin{document}

\begin{center}
{\Large 
{\bf On the Mechanism of Temperature Variations in the Average Energy of Muons at Large Depths}
}

\vskip 0.5cm
{\bf N.~Yu.~Agafonova and A.~S.~Malgin}

~~~
{\it  Institute for Nuclear Research of the Russian Academy of Sciences, Moscow, Russia}
\end{center} 

\noindent E-mail: Agafonova@inr.ru

\noindent Date: 11 August 2020

\begin{abstract}
Sources of seasonal temperature variations in the average energy of the muon flux detected in the
LVD experiment have been discussed. It has been shown that variations are due to the processes of generation of muons in upper layers of the atmosphere and the passage of muons through a thick rock layer.

\end{abstract}


\vskip 0.5cm
\noindent
{\bf
A full version was published in 
Journal of Experimental and Theoretical Physics, 2021, Vol. 132, No. 1, pp. 73-78.
}

\noindent  Zhurnal Eksperimental’noi i Teoreticheskoi Fiziki, 2021, Vol. 159, No. 1, pp. 88–94

\vskip 1.5cm

\section{Introduction}
Seasonal variations in atmospheric muons deep underground are sill of interest mainly because they
contain information, first, on cyclic processes in upper layers of the atmosphere and, second, on the
characteristics and time behaviour of the background in low-background underground experiments.

We focus on the second aspect of studies of variations in the muon flux. A reason for seasonal 
variations in muons at large depths is a positive temperature effect responsible for change in the density of the atmosphere and its height caused by heating in summer and cooling in winter. 
The amplitude ${\delta I_{\mu} = 1.5\%}$ and the phase of seasonal variations in the muon
intensity ${\phi(I_{\mu}) = (185 \pm 15)}$ day at a depth of ${\sim}$ 3.6 km w.e. were determined in the experiments \cite{1},\cite{2},\cite{3},\cite{4},\cite{5},\cite{6},\cite{7},\cite{8},\cite{9}. In addition, the amplitude of seasonal variations in
neutrons ${\delta \Phi_n}$ formed by a muon flux was determined in the experiment reported in \cite{10}. The quantity ${\delta \Phi_n}$ surprisingly appeared to be larger than the amplitude
$\delta I_{\mu}$ by about a factor of 6. It was previously accepted {\it a priori} that variations
${\delta \Phi_n}$ should be equal to variations in the muon flux ${\delta I_{\mu}}$. It was shown in \cite{11} that the detected effect can be explained by variations in the average energy of muons 
${\delta \overline{E}_{\mu}}$. According to the experimental data reported in \cite{10}, the amplitude of variations ${\delta \overline{E}_{\mu}}$ at the depth of the LVD experiment should be approximately 10\% in order to ensure, together with variations in the muon intensity ${\delta I_{\mu}=1.5\%}$, the amplitude of the variations in the cosmogenic neutron flux ${\delta \Phi_n=9.3\%}$ measured in the LVD experiment \cite{10}.

In this work, we consider processes responsible for seasonal variations in the average energy
${\overline{E}_{\mu}}$ of the muon flux. Section 2 presents the general expressions
characterizing the relation of the spectrum of muons at the depth of the LVD experiment to the spectrum of
muons at the sea level (sl). In Section 3, we determine the energy spectrum of the LVD-muon flux and their seasonal variation underground and at the sea level. The processes of generation of high-energy muons responsible for seasonal variations in the muon flux are discussed in Section 4. In Sections 5 and 6, the LVD and Borexino results are compared and the accuracy of determination of ${\delta \overline{E}_{\mu}}$ is estimated.

\section{Relation Between the Energy Characteristics of LVD Muons Underground and at the Sea Level}

Temperature variations in the muon intensity are due to the processes of their generation in upper layers of the atmosphere. The subsequent passage of muons through the atmosphere hardly affects the energy of LVD muons, i.e., muons reaching the LVD depth ($H^{min} = 3.1$ km w.e.). We consider the LVD data because they exhibit anomalous variations in cosmogenic neutrons.

The energy losses of muons in rock transform their initial energy spectrum on the ground (below, at the
sea level) but do not change the intensity of muons whose energy at the sea level is not below the threshold value $E^{th}_{\mu}$, i.e., enough to reach the LVD depth. Consequently, the mechanism of variations in the average energy of muons $E^{th}_{\mu}$ underground should involve processes of both generation of muons in upper layers of the atmosphere and, in contrast to variations in the intensity, passage of muons through a thick rock layer.

To determine sources of variations, we consider the relation of the characteristics (intensity, effective
energy range, average energy) of LVD muons at the depth $H^{min}$ to their characteristics on the ground and, then, to the characteristics of “parent” pions and generation of pions in $pA$-collisions. The analysis will involve single muons because they constitute $90\%$ of the total number of muons ($10\%$ of muons enter muon groups) reaching the LVD depth \cite{12}. The average energy of single LVD muons 
${\overline{E}_{\mu}} = (270 \pm 18)$ GeV was obtained in the measurements reported in \cite{13}. Thus, ${\overline{E}_{\mu}} = 270$ GeV is accepted in the analysis.

The minimum energy of muons at the sea level necessary for reaching the depth $H^{min} = 3.1$ km w.e. is
${{E}^{min}_{\mu,\mathrm{sl}}} = 1.3$ TeV \cite{14}. The threshold energy ${{E}^{th}_{\mu,\mathrm{sl}}}$(50\% probability of survival) is 1.8 TeV. This value can be determined using the following relation of the energy of a muon at the sea level $E_{\mu,\mathrm{sl}}$ to its average energy $E_{{\mu},H}^{\mathrm{av}}$ at the depth $H$ \cite{15}:
\begin{equation}
E_{{\mu},H}^{\mathrm{av}}=(E_{{\mu},\mathrm{sl}}+\epsilon_{\mu})e^{-bH}-\epsilon_{\mu}
\end{equation}
Therefore,
\begin{equation}
E_{{\mu},\mathrm{sl}}=(E_{{\mu},\mathrm{sl}}^{\mathrm{av}}+\epsilon_{\mu})e^{+bH}-\epsilon_{\mu}
\end{equation}
Setting $E_{{\mu},H}^{\mathrm{av}} = 0$, for $E_{{\mu},\mathrm{sl}}^{th}$ we get:
\begin{equation}
E_{{\mu},\mathrm{sl}}^{th}=(e^{+bH}-1)\epsilon_{\mu}
\end{equation}
Here, $\epsilon_{\mu}$=$a/b$ is the parameter characterizing the shape of the differential spectrum of muons, which is quasi-flat at large depths:
$\frac{dN_{\mu}}{dE} \propto 1/(\epsilon_{\mu}+E_{\mu})^{\gamma_{\mu}}$.\\
The parameter $\epsilon_{\mu}$ is an energy above which radiation losses begin to dominate and the quasi-flat spectrum of muons acquires the form of a spectrum on the ground 
$P_{\mathrm{sl}}(E_{\mu})\propto E_{\mu}^{-\gamma_{\mu}}$. The energy losses of muons in the matter layer with the thickness $H$ is given by the expression
\begin{equation}
-dE_{\mu}/dH=a+bE_{\mu}
\end{equation}
Here, $a$ are the total ionization losses and $b$ are the total losses in three radiative processes. The ratio $a/b = \epsilon_{\mu}$ is the critical energy of a muon at which ionization losses are equal to radiation losses; ionization losses dominate at $E_{\mu} \ll \epsilon_{\mu}$, whereas radiation losses
prevail at $E_{\mu} \gg \epsilon_{\mu}$ The parameters $a$ and $b$ depend slightly on $E_{\mu}$: as $E_{\mu}$ 
increases from $1$ TeV to $10$ TeV, the parameter $a$ in the standard rock increases from $268$ to $293$ 
GeV(km w.e.)$^{-1}$ and the parameter $b$ increases from $0.392$ to $0.435$ (km w.e.)$^{-1}$ 
(see Table 24.2 in \cite{15}).

The substitution of the parameter $\epsilon_{\mu} = a/b = 667$ GeV ($a$ = 280 GeV/km m.e., 
$b$ = 0.42/km w.e.) for the depth $H$ = 3.1 km w.e. into Eq. (3) gives $E^{th}_{\mu,\mathrm{sl}}= 1785$ GeV $\approx$ 1.8 TeV. 
With the chosen parameters $a$ and $b$, the average energy of single LVD muons $\overline{E}_{\mu}^{cal}$ = 277 GeV, calculated by the formula from ref. \cite{16}
\begin{equation}
\overline{E}_{\mu}^{cal}=\epsilon_{\mu}[1-exp(-bH)](\gamma_{\mu}-2)^{-1},
\end{equation} 
agrees with high accuracy with the experimental value $\overline{E}_{\mu} = 270$ GeV.

The parameter $\gamma_{\mu}=3.75$  is the absolute value of the exponent of the differential spectrum of muons at the sea level: $P_{\mathrm{sl}}(E_{\mu}) \propto E_{\mu}^{-\gamma_{\mu}}$.

Here we can note an interesting property of the $\overline{E}_{\mu}$ value: it is “saturated.” 
At large depths ($H \gg 1/b$), in formula (5) for $\overline{E}_{\mu}^{cal}$, the factor $[1 - exp(-bH)]$ is approximately $1$, which at $\epsilon_{\mu}$ = const leads to the expression of the limiting average energy of the single atmospheric muon flux in the standard rock:
\begin{equation}
\overline{E}_{\mu}^{lim} = \epsilon_{\mu}(\gamma_{\mu}-2)^{-1}.
\end{equation}
The average muon energy $\overline{E}_{\mu}$ at depths $\geq5$ km w.e. asymptotically approaches the energy $\overline{E}_{\mu}^{lim}$. The limit energy at the parameters 
       $\epsilon_{\mu}=693$ GeV and $\gamma_{\mu}=3.77$ used in \cite{16} is $\overline{E}_{\mu}^{lim}=392$ GeV; 
    at $\epsilon_{\mu}=618$ GeV and $\gamma_{\mu} = 3.7$, $\overline{E}_{\mu}^{lim}=364$ GeV \cite{17}; 
and at $\epsilon_{\mu}=495$ GeV and $\gamma_{\mu} = 3.7$, $\overline{E}_{\mu}^{lim}=291$ GeV \cite{18}. 
As seen, the parameter is calculated with a large error. Taking into account various measurements, 
we can accept $\overline{E}_{\mu}^{lim}$ = 400 GeV.

\section{Seasonal Variations in the Energy Spectrum of LVD Muons Underground and at the Sea Level}

The differential spectrum of muons at depths $H > 1/b \approx 2.5$ km w.e. is quasi-flat to the energy
$\sim \epsilon_{\mu}$, above which the spectrum becomes steeper, acquiring the form $P_{\mathrm{sl}}(E_{\mu}) \propto E_{\mu}^{-\gamma_{\mu}}$  with the exponent $\gamma_{\mu}$ = 3.75. Consequently, the spectrum of LVD muons can be represented in the form of a step cut at the energy
$^0E_{\mu}^{max}$ = 2$\overline{E}_{\mu}$ = 2$\cdot$270 GeV = 540 GeV (stepwise spectrum
approximation), $^0E_{\mu}^{max}$ is the annual average energy of the “step.” In this case, the energies of all LVD muons at a depth of 3.1 km w.e. with the intensity $^0I_{{\mu},H}$ are in the range of $0-540$ GeV.

Substituting $\overline{E}_{{\mu},H}^{\mathrm{av}}$  = $^0E_{\mu}^{max}$ = 540 GeV, into Eq. (2),
we determine the corresponding energy at the sea level: $^0E_{{\mu},\mathrm{sl}}^{max}$  = 3771 GeV $\approx$ 3.8 TeV. Therefore, the annual average spectrum of muons at the LVD depth is effectively formed by LVD muons whose energy at the sea level is in the range from $E_{{\mu},\mathrm{sl}}^{th}$ to $^0E_{{\mu},\mathrm{sl}}^{max}$ , i.e., in the range of $ 1.8 - 3.8$ TeV, with the intensity $^0I_{\mu,\mathrm{sl}}$ = $^0I_{\mu,H}$.

Within the hypothesis of the relation of the seasonal variations in the number of cosmogenic neutrons to the average energy of muons, which explains anomalous variations in the number of neutrons, the
energy of muons $\overline{E}_{\mu}$ at the LVD depth in the summer period $^s\overline{E}_{{\mu},H}$
increases by $10\%$. It is noteworthy that the parameter $\overline{E}_{\mu}$ is a natural energy parameter  characterizing the flux of both muons and neutrons formed by them although most of the neutrons are produced by high-energy muons.

At the stepwise spectrum of muons, an increase in $^s\overline{E}_{{\mu},H}$ by $10\%$ 
should increase the maximum energy of the spectrum also by $10\%$:
$^sE_{{\mu},H}^{max}$ $= 1.1 \cdot $ $^0E_{{\mu},H}^{max}$ = 594 GeV. According to Eq. (2), this value corresponds to the energy at the sea level $^sE_{{\mu},\mathrm{sl}}^{max} = 3970$ GeV. Thus,
an increase in $E_{{\mu},H}^{max}$ by a factor of $1.1$ (from $^0E_{{\mu},\mathrm{sl}}^{max}$ = 3771 GeV to $^sE_{{\mu},\mathrm{sl}}^{max}$ = 3970 GeV)  is due to an increase in the energy 
$^0E_{{\mu},\mathrm{sl}}^{max}$ by $5.3\%$. At the same time, an
increase in $\overline{E}_{{\mu},H}$= 270 GeV by a factor of $1.1$ corresponds to an increase in the energy $E_{\mu,\mathrm{sl}}$ only by a factor of $1.036$ (from 2780 to 2880 GeV). The resulting
relations are explained by the property of Eq.(2), which relates the energies $E_{\mu,H}^{av}$ and $E_{\mu,\mathrm{sl}}$ and reflects the influence of quasi-static ionization losses and radiation energy losses on the shape of the underground muon spectrum. However, the fluctuation character of radiation losses, on one hand, significantly increases the probability that muons reach large depths and, on the other hand, nonlinearly “expands” the spectrum of muons deep underground. Because of these losses, the monoenergetic flux of muons with the energy at the sea level, e.g., $10$ TeV is transformed at various depths to spectra of various shapes (see Fig. 1 in \cite{19}).

\begin{figure}
%\center{\includegraphics[width=1\linewidth]{ffig01}}
\center{\includegraphics[width=12cm]{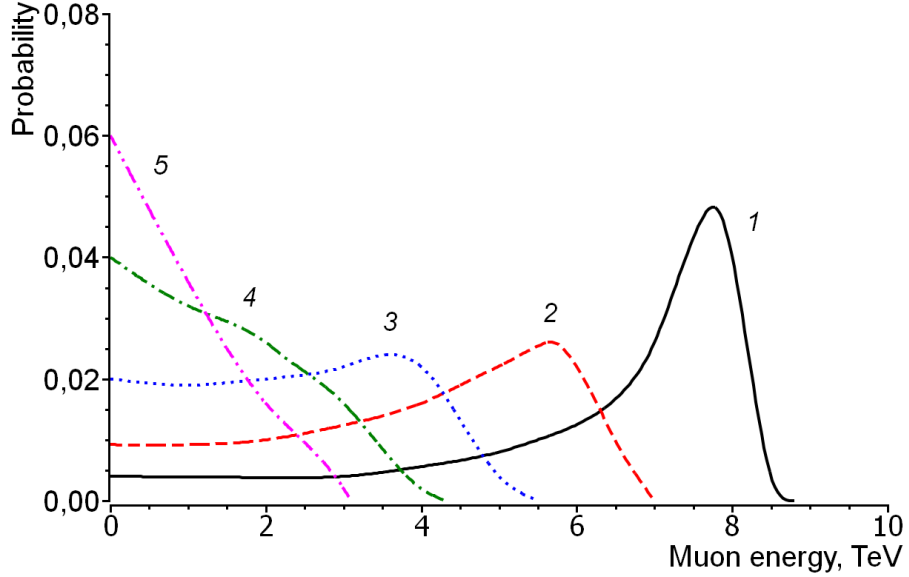}}
\caption{Probability distribution for a muon with an energy of 10 TeV at the sea level to have the energy from $E$ to $E + 0.1$ TeV at depths $5$, $4$, $3$, $2$, and $1$ km w.e. (indicated near the corresponding lines).}
\end{figure}

The parameters $^0E_{{\mu},\mathrm{sl}}^{max}$ and $^sE_{{\mu},\mathrm{sl}}^{max}$ are really determined by the segments of the spectrum of muons $P_{\mathrm{sl}}(E_{\mu})$ whose energies exceed these values. These values found above in the stepwise approximation for the LVD muon spectrum are close to real because the spectrum decreases rapidly as $P_{\mathrm{sl}}(E_{\mu})\propto E_{\mu}^{-3.75}$.

\section{Seasonal Variations in the Energy Range of Effective Generation of LVD Muons}

The average energy of the muon is related to the energy of the “parent” pion as $E_{\pi}=(m_{\pi}/m_{\mu})E_{\mu}$, where $m_{\pi}$ and $m_{\mu}$ are the masses of the pion and muon, respectively. Using this relation and neglecting ionization energy losses of muons in air ($\approx 2$ GeV), one can
pass from the energy $E_{\mu,\mathrm{sl}}^{th}$ to the threshold energy of the pion: 
$E_{\pi}^{th}\approx(m_{\pi}/m_{\mu})E_{\mu}^{th}$ = 2.4 TeV. Let only about
$5\%$ of 2.4-TeV pions decay (below, we will show that $k_{\pi}^{dec}$ = 0.05 is due to the dominance of single muons in the total flux at the LVD depth) and the other pions generate secondary hadrons. Under these assumptions, for the production of a single muon with the energy $E_{\mu}^{th} = 1.8$ TeV by pions of the first generation, it is necessary to spend an energy of about 48 TeV on the production of charged pions. The sum of this energy and the energy of neutral pions (about half the energy of $\pi^{\pm}$) is the interaction energy $E_{in}^{th}\approx$ 72 TeV with the production of pions. The average inelasticity
factor $K_{inl}$ in deep inelastic $pA$-collision is approximately $0.5$ (about half the energy $E_p$ is carried by the leading nucleon). Correspondingly, the threshold energy of the proton necessary for the production of a muon reaching the LVD depth is $E_p^{th} \approx$ 144 TeV. The rapidly decreasing spectrum of primary protons $F(E_p) \propto E_p^{-2.75}$ and the fast increase in the atmosphere density with decreasing altitude (which increases the probability of ${\pi}A$-inelastic collisions for pions of the second generation) are responsible for the decisive role of pions of the first generation in the formation of the muon flux at the LVD depth. The contribution from pions of the second and next generations to the
generation of high-energy muons does not exceed $20\%$ \cite{20}.

Repeating the same calculations for the annual
average $^0E^{max}$ and summer $^sE^{max}$ values, we obtain $^0E_{\pi}^{max}= 4.98$ TeV and $^sE_{\pi}^{max}= 5.24$ TeV. Consequently, the segments of the effective generation of LVD muons in pion spectra are limited by the energies  $2.4 \leq$ $^0E_{\pi} \leq 4.98$ TeV, $2.4 \leq$ $^sE_{\pi} \leq 5.24$ TeV.

Using the dependence of the multiplicity of pions
$\nu_{\pi} \approx 3\cdot ln E_{in}$ on the interaction energy $E_{in}$ and assuming
that $k_{\pi}^{dec} = 0.05$ independent on $E_{in}$, one can estimate the number of muons $N_{\mu}$ that are produced at the threshold, $E_{in}^{th} = 72$ TeV, and the maximum, $E_{in}^{max} = 152$ TeV, interaction energies and reach the LVD depth:
$\nu_{\pi}^{\pm} = 2/3 (3 lnE_{in})$, 
$N_{\mu} = k_{\pi}^{dec}\cdot\nu_{\pi} = 0.1\cdot lnE_{in}$ [GeV]; therefore, 
$N_{\mu}^{th} = 1.12$ and $N_{\mu}^{max} = 1.19$. Thus, one $pA$-collision event in the energy range
 72 $\leq E_{in} \leq$ 152 TeV corresponds to one LVD muon.

To efficiently generate LVD muons in decays of pions, the energy $E_{\pi}$ should be below the critical
energy $E_{\pi}^{cr}$ corresponding to the density of the atmosphere $\rho_{at}$ at the altitude of the production of pions of the first generation. The energy $E_{\pi}^{cr}$ is determined by the equality 
$\lambda_{\pi}^{dec} = \lambda_{\pi}^{in}$,
where \\
$\lambda_{\pi}^{dec} = \gamma_{f}\tau_0c_0=(E_{\pi}/m_{\pi}c_0^2)\tau_0c_0$ 
is the decay mean free path of a relativistic pion, \\
$\tau_0 = 2.6\times10^{-8}$ s is the lifetime of the pion at rest,\\ 
$\gamma_f$  is the gamma factor of the pion, \\
$c_0 = 3\times10^{10}$ cm/s,\\
$\lambda_{\pi}^{in}$ = $(\sigma_{\pi}^{in}n_A\rho_{at})^{-1}$ is the mean free path of
the pion for the inelastic $\pi A$-interaction, \\
$n_A$ is the number of nuclei in gram of air, 
and\\
$\sigma_{\pi}^{in}$ is the cross section for the inelastic $\pi A$-interaction. 
For the relativistic pion in air, 
$(\sigma_{\pi}^{in}n_A)^{-1}$ = 120 g/cm$^2$ and $\lambda_{\pi}^{in}$ = 120$\rho_{at}^{-1}$. \\
Consequently, $E_{\pi}^{cr}$ [GeV] $=120\rho_{at}^{-1}(m_{\pi}c^2/\tau_0c_0)=2.15\times10^{-2}\rho_{at}^{-1}$.
Setting $E_{\pi}^{th} = E_{\pi}^{cr} = 2.4$ TeV, one can estimate the density of the air layer where pions are generated as $\rho_{at} \approx 9\times10^{-6}$ g/cm$^3$, which corresponds to the density of
the atmosphere at an altitude of about 35 km.

Thus, most of the LVD muons are produced in decays of pions of the first generation with energies in
the range from $E_{\pi}^{th}$ to $E_{\pi}^{cr}$. The energy $^0E_{\pi}^{cr}$ corresponds to the annual average altitude of the atmospheric layer, where pions with the energy $E_{\pi}\geq E_{\pi}^{th}$ are
efficiently generated. The summer increase in the atmospheric temperature results in an increase in the
altitude and expansion of the generation layer of pions with the energies $E_{\pi}\geq E_{\pi}^{th}$, which is accompanied by a decrease in the density $\rho_{at}$. Because of the decrease in the density of the boundary atmospheric layer at an altitude of about 40 km, the energy range of decaying pions is expanded owing to an increase in the energy $E_{\pi}^{cr}$ from $^0E_{\pi}^{max}$ = 4.98 TeV to the summer value $^sE_{\pi}^{cr}$ = $^sE_{\pi}^{max}$ = 5.24 TeV and the spectrum of produced
muons is hardened.

The summer hardening of the spectrum of LVD
muons at the sea level ($^s\gamma_{\mu,\mathrm{sl}}$ $<$ $^0\gamma_{\mu,\mathrm{sl}}$ = 3.75), which is responsible for the increase in the energy $\overline{E}_{\mu,\mathrm{sl}}$ and intensity $I_{\mu}$, has the same reason as the hardening of the energy spectrum and the increase in the intensity of high-energy muons ($\geq 1$ TeV) with increasing observation angle $\theta$. This reason is an increase in the mean free path of pions in the low-density atmospheric layer, which increases the probability of the decay of high-energy pions. In particular, the measurements reported in \cite{21} show that the differential spectrum of muons at the sea level averaged over the angular interval 
$55^{\circ} < \theta < 90^{\circ}$ in the energy range $2\times10^2$ GeV $< E_{\mu}< 3\times10^3$ GeV is described by a power law with the exponent $\gamma_{\mu}$ = 3.1 (error of $7\%$).

The summer deviation of the parameter $^s\gamma_{\mu,\mathrm{sl}}$ from the annual average value $^0\gamma_{\mu,\mathrm{sl}}$ = 3.75 can be estimated using Eq. (5) and the $10\%$ amplitude of variations in the energy at the LVD depth. Assuming that the numerator in Eq. (5) is almost constant during the
year, we obtain
\begin{equation}
\frac{^s\overline{E}_{\mu,LVD}}{^0\overline{E}_{\mu,LVD}}=\frac{^0\gamma_{\mu,\mathrm{sl}}-2}{^s\gamma_{\mu,\mathrm{sl}}-2}=1.1.
\end{equation}
According to Eq. (7), $^s\gamma_{\mu,\mathrm{sl}}$ = 3.59 at $^0\gamma_{\mu,\mathrm{sl}}$ = 3.75;
i.e., the exponent $\gamma_{\mu,\mathrm{sl}}$ decreases in summer by $4.3\%$,
which is within the accuracy of measurements of the parameter $\gamma_{\mu,\mathrm{sl}}$ and complicates the determination of variations in the exponent $\gamma_{\mu,\mathrm{sl}}$ in the experiment.

The summer change in the shape of the underground muon spectrum (Fig. 2) is similar to the transformation of the spectrum with increasing depth. However, an increase in the depth results in an increase in 
$\overline{E}_{\mu}$ and a decrease in the intensity $I_{\mu}$, whereas the summer change in the shape of the spectrum $P_H(E_{\mu})$ at the depth $H$ leads to an increase in both the energy $\overline{E}_{\mu}$ and intensity $I_{\mu}$.

Since the annual average characteristics of the muon flux are constant, their change in summer is opposite to the change in winter.

\begin{figure}
\center{\includegraphics[width=10cm]{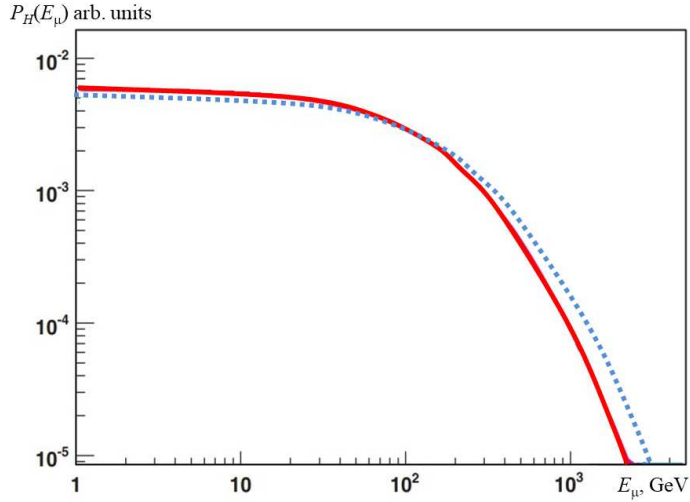}}
\caption{Qualitative representation of the seasonal transformation of the shape of the muon spectrum. The red solid line is the annual average spectrum $^0P_H(E_{\mu})$ and the blue dashed line is the summer spectrum $^sP_H(E_{\mu})$.}
\end{figure}

\section{Discussion}

Variations in cosmogenic neutrons make it possible to detect seasonal variations in the average energy of
muons at large depths, where $\overline{E}_{\mu}\geq$ 200 GeV. This method is based on the dependence of the neutron yield on the energy of muons $Y_n \propto \overline{E}_{\mu}^{0.78}$, which is confirmed in numerous experiments and is supported phenomenologically \cite{22}.

Necessary conditions for the determination of variations $\delta\overline{E}_{\mu}$ are a sufficient count rate of muons, stable long-term (no less than several years) operation of a setup, and a high efficiency of detection of neutrons. 
Variations in the mean energy of underground muons $\delta\overline{E}_{\mu}$
 can be detected by the direct measurement of the energy of muons by the TRD \cite{13} and parameter \cite{23} methods. However, these methods are low efficient for the determination of variations $\delta\overline{E}_{\mu}$ because they cannot ensure the necessary methodological conditions.

Seasonal variations in the energy $\overline{E}_{\mu}$ were revealed not only at the LVD, but also in the long-term Borexino experiment \cite{24}, performed near the LVD. The amplitude of variations in the production rate of cosmogenic neutrons and the corresponding amplitude of variations $\delta\overline{E}_{\mu}$ in the Borexino experiment are $2.6\%$ and 9.2 GeV, respectively, which are one third of the
respective parameters obtained at the LVD ($7.7\%$ and 28 GeV). The difference between the LVD and Borexino results is most probably due to the features of the methods of determining the parameter $N_n$ and data processing.

The selection of muon events with the multiplicity of neutrons below 10 for the analysis can possibly
affect the Borexino result. Although the events with the multiplicities $\geq$10 constitute only a small fraction of the total number of muons, these events most strongly affect variations in the underground muon spectrum because they are related to the high-energy region of the muon spectrum underground in view of the dependence $N_n \propto  \overline{E}_{\mu}^{0.78}$.

Experiments on the detection of seasonal variations in the energy of atmospheric muons by direct measurements of the muon energy at the sea level, as well as underground experiments, should satisfy the requirement of the long-term operation with stable parameters and have a sufficient energy resolution.
The existing measurements of the energy spectrum of muons at the sea level in the energy range $E_{\mu} > 1$ TeV (Fig. 4 in \cite{25}) show that the performed experiments do not satisfy this requirement: the spread of data from different experiments is much larger than the error of measurements indicated on the plot. Furthermore, the research programs did not include the search for seasonal variations in the average energy of the muon flux at the sea level at energies above 1 TeV.

\section{Conclusion}

A larger amplitude of seasonal variations in cosmogenic neutrons compared to the amplitude of variations
in the muon intensity has been detected in two underground experiments. Variations in the number of neutrons are related to variations in the average energy of muons as $N_n \propto  \overline{E}_{\mu}^{0.78}$. Consequently, the temperature effect affecting the generation of muons changes not only their intensity but also the average energy. Seasonal variations in the average energy of atmospheric muons constitute a new effect in muon physics.

The stepwise spectrum approximation for LVD muons underground makes it possible to pass from variations in the average energy of muons underground to variations in the energy of the step $E_{\mu}^{max}$ when studying the mechanism of variations.

The consideration of variations in the energy range of muons in the path from the generation layer in the
atmosphere to a depth of 3.1 km w.e. in the stepwise spectrum approximation for LVD muons underground indicates that variations $\delta\overline{E}_{\mu}$ are determined by variations in the critical energy for pions $E_{\pi}^{cr}$. As a result, the range of generation of muons $E_{\mu,\mathrm{sl}}^{th}-E_{\mu,\mathrm{sl}}^{max}$ is varied with an amplitude of $5.3\%$ with the subsequent increase in variations to 
$\delta E_{\mu}^{max}$ = $\delta\overline{E}_{\mu}$ = $10\%$ after the passage of muons through the 3.1-km w.e. rock layer.

Seasonal variations in the generation of high-energy muons at the edge of the atmosphere are also manifested in variations of the hardness of their spectrum at the sea level and in the variation of the shape of quasi-step spectrum of muons underground.

The accuracy of determining variations in generation of muons responsible for variations $\delta\overline{E}_{\mu}$ = $10\%$ depends on the consistency of the stepwise muon spectrum approximation with the characteristics of the real spectrum underground, as well as on errors of the calculations of energy ranges by Eqs. (1)–(3) and (5) and parameters entering into them. Taking into account uncertainty of these parameters (on the example of the calculation of the energy $\overline{E}_{\mu}^{lim}$), it can be expected that the accuracy of determining variations $\delta\overline{E}_{\mu}$ is no worse than $20\%$.

The complete simulation of variations $\delta I_{\mu}$ and $\delta\overline{E}_{\mu}$ (including atmospheric phenomena, transformation of the spectrum of high-energy muons in the rock, and generation of neutrons) still gives contradictory results inconsistent with experimental data \cite{2}, \cite{26}.

Since seasonal variations in the neutron flux are noticeable, they should be taken into account when measuring the neutron yield. The temperature of the atmosphere undergoes not only seasonal modulations but also irregular variations during the year. As a result, the number of neutrons produced by muons underground significantly deviates from the annual average value and from a harmonic function, having a nonconstant amplitude of modulations and a nonconstant phase of oscillations. This should be taken into account when analyzing the background in low-background underground experiments.

\textit{Acknowledgments}. This work was supported by the Russian Foundation for Basic Research (project no. 18-02-00064-a) and by the program of international cooperation between INFN (Italy) and the Ministry of Science and Higher Education of the Russian Federation.

\end{document}